\documentclass[aps,prd,10pt,preprint,superscriptaddress,onecolumn,nofootinbib]{revtex4-1}
\usepackage{graphicx, epsfig} 
\usepackage{amsmath,amssymb,amsfonts,dsfont,mathrsfs,amsthm,mathtools}
\usepackage{bm} 
\usepackage{color}
\usepackage[usenames]{xcolor}
\usepackage{hyperref}
\usepackage{siunitx}
\hypersetup{colorlinks=true,urlcolor=blue,linkcolor=magenta,citecolor=blue,filecolor=blue}
\usepackage[normalem]{ulem}
\usepackage{array}
\usepackage{hyperref}
\usepackage{booktabs}
\usepackage{blindtext}

\newcommand{\diff}[1]{\text{d}#1}
\newcommand{\Diff}[1]{\text{D}#1}

\begin{document}
	
\title{Asymptotically AdS black hole with a conformally-coupled scalar field in the first-order formalism of gravity}

	\author{Luis Avilés}
	\email{luaviles@unap.cl}
	\affiliation{Instituto de Ciencias Exactas y Naturales, Universidad Arturo Prat, Playa Brava 3256, 1111346, Iquique, Chile}
	\affiliation{Facultad de Ciencias, Universidad Arturo Prat, Avenida Arturo Prat Chac\'on 2120, 1110939, Iquique, Chile}

	\author{Crist\'obal Corral}
\email{cristobal.corral@uai.cl}
\affiliation{Departamento de Ciencias, Facultad de Artes Liberales, Universidad Adolfo Ibáñez, Avda. Padre Hurtado 750, 2562340, Viña del Mar, Chile}

	\author{Fernando Izaurieta}
	\email{fizaurie@udec.cl}
	\affiliation{Faculty of Engineering, Architecture and Design, Universidad San Sebasti\'an, Lientur 1457, Concepci\'on, Chile.}
	\author{Omar Valdivia}
	\email{ovaldivi@unap.cl}
	\affiliation{Instituto de Ciencias Exactas y Naturales, Universidad Arturo Prat, Playa Brava 3256, 1111346, Iquique, Chile}
	\affiliation{Facultad de Ciencias, Universidad Arturo Prat, Avenida Arturo Prat Chac\'on 2120, 1110939, Iquique, Chile}
	
	\author{Carlos Vera}
	\email{carlosvera02@unap.cl}
	\affiliation{Instituto de Ciencias Exactas y Naturales, Universidad Arturo Prat, Playa Brava 3256, 1111346, Iquique, Chile}
	\affiliation{Facultad de Ciencias, Universidad Arturo Prat, Avenida Arturo Prat Chac\'on 2120, 1110939, Iquique, Chile}

\begin{abstract}

We present a novel asymptotically anti-de Sitter black hole solution with conformally-coupled scalar fields in the first-order formalism of gravity in four dimensions. To do so, we consider a one-parameter extension of conformal transformations by exploiting the fact that the tetrad and spin connection are regarded as independent fields. We solve the field equations analytically and obtain a static black hole solution with nontrivial torsion sourced by the conformal coupling between the scalar field and geometry. The presence of torsion renders the scalar field everywhere regular, while the curvature and torsion singularities coalesce into the origin. We show that this configuration is continuously connected to previously reported solutions in the limit of vanishing torsion and analyze its main properties, focusing on the consequences of the torsional singularity.     
\end{abstract}

	\maketitle

	\section{Introduction}

Black holes are one of the most fascinating objects in nature. They were predicted by Einstein's theory of general relativity, and their existence has been well established through different experimental evidence~\cite{Kormendy:1995er,Celotti:1999tg,Gillessen:2008qv,Mu_oz_2011,LIGOScientific:2016aoc,EventHorizonTelescope:2022wkp}. In electrovacuum, they are uniquely characterized by their mass, angular momentum, and electric charge, that is, the physical quantities that can be measured as asymptotic charges at spatial infinity. Any other type of imprint that matter fields could leave on black holes is lost once its stationary phase is achieved; this leads to a uniqueness theorem for the Kerr-Newman black hole in Einstein-Maxwell theory.\footnote{This can be circumvented in Einstein-Yang-Mills theory~\cite{Volkov:1989fi,Bizon:1990sr,Kuenzle:1990is,Baltsov:1991au} and generalized Proca theories~\cite{Heisenberg:2017hwb}.} This is known as the no-hair theorem~\cite{Israel:1967wq,Israel:1967za,Carter:1971zc,Ruffini:1971bza,Bekenstein:1995un,Chrusciel:2012jk}, where hair is referred to as any other asymptotic charge apart from the aforementioned quantities that can be measured by a Gauss law at infinity.  

The no-hair theorem is a theory-dependent statement and it relies critically on the minimal coupling of matter to gravity. For instance, in asymptotically flat spacetimes, it is known that minimally-coupled scalars lead to a solution that is not a black hole but, rather, represents a naked singularity~\cite{Janis:1968zz,Janis:1969ivo}. One could, in turn, allow for nonminimal-scalar coupling to circumvent this issue. A natural possibility would be to consider conformally-coupled scalar fields. Nevertheless, the static and spherically symmetric solution found by Bekenstein~\cite{Bekenstein:1974sf} and independently by Bocharova-Bronnikov-Melnikov~\cite{Bocharova:1970skc} reveals that this class of coupling leads to a scalar field that becomes singular at the horizon. Even though it was argued that this singularity is not necessarily problematic from the viewpoint of an asymptotic observer~\cite{Bekenstein:1975ts}, it was shown that it renders an ill-defined stress-energy tensor, whose evaluation through suitable regularization methods yields ambiguous results~\cite{Sudarsky:1997te}. Remarkably, this problem can be solved by introducing a cosmological constant and a conformal potential for the scalar field, pushing the scalar-field singularity behind the event horizon~\cite{Martinez:2002ru,Martinez:2005di}. After this resolution was proposed, different configurations with conformally-coupled scalar fields have been found in four dimensions~\cite{Anabalon:2009qt,Charmousis:2009cm,Astorino:2014mda,Anabalon:2012tu,Bardoux:2012tr,Astorino:2013sfa,Bardoux:2013swa,Cisterna:2021xxq,Barrientos:2022avi,Barrientos:2022yoz,Babichev:2022awg,Barrientos:2023tqb}.\footnote{Solutions with scalar hair have been found in Horndeski gravity as well~\cite{Rinaldi:2012vy,Anabalon:2013oea,Bravo-Gaete:2013dca,Sotiriou:2014pfa,Cisterna:2014nua,Bravo-Gaete:2014haa,Cisterna:2015yla,Brihaye:2016lin,Cisterna:2016vdx,Babichev:2016rlq,Jiang:2017imk,Arratia:2020hoy}.} In higher dimensions, however, the standard conformal coupling does not admit black hole solutions~\cite{Xanthopoulos:1992fm,Klimcik:1993cia} and one needs to introduce conformal higher-curvature corrections to circumvent this no-go theorem~\cite{Giribet:2014bva,Chernicoff:2016jsu,Babichev:2023rhn}. 

A natural question is whether relaxing some assumptions on the geometry could help to find a way out to the no-hair theorem. In the first-order formalism of gravity, i.e. where the metric and connection are treated as independent dynamical fields, it is known that nonminimally coupled scalar fields act as a source of torsion (See~\cite{Barrientos:2017utp} and references therein). This class of interaction appears in different contexts. For instance, the dimensional reduction of low-energy limits of string theory, e.g. Einstein-Gauss-Bonnet gravity~\cite{Zwiebach:1985uq}, induces a nonminimal coupling between the dilaton and topological terms of the Euler class~\cite{Charmousis:2014mia,Lu:2020iav,Hennigar:2020lsl,Ma:2020ufk}. In the context of cosmology, the torsion-induced interaction might explain the accelerated expansion of the Universe from a geometrical viewpoint~\cite{Toloza:2013wi,Espiro:2014uda,Castillo-Felisola:2016kpe,Cid:2017wtf}. On the other hand, the shift symmetry in the linear pseudo-scalar-Nieh-Yan coupling~\cite{Hojman:1980kv,Nelson:1980ph,Nieh:1981ww} allows one to renormalize the divergent piece coming from the torsional contribution to the axial anomaly~\cite{Chandia:1997hu,Mercuri:2009zi}. The latter generates an axion-like particle that could provide a solution to the strong CP problem from a gravitational perspective~\cite{Lattanzi:2009mg,Castillo-Felisola:2015ema}. In this context, axionic hair on slowly-rotating black holes~\cite{Campbell:1990ai,Duncan:1992vz} and locally $\text{AdS}_{3}\times\mathbb{R}$ black string solutions have been found~\cite{Cisterna:2018jsx,Corral:2021tww}; as well as in the first-order formulation of Chern-Simons modified gravity~\cite{Jackiw:2003pm,Alexander:2008wi,Alexander:2009tp,Yunes:2009hc,Konno:2009kg,Amarilla:2010zq,Nashed:2023qjm}. Finally, from a phenomenological viewpoint, (pseudo-)scalar-induced torsion leads to luminal propagation of gravitational waves, although their polarization can experience birefringence in vacuum~\cite{Barrientos:2017utp,Elizalde:2022vvc,Sulantay:2022sag}. All of this suggests that torsional scalar-tensor theories could provide a novel way to circumvent the no-hair theorem by relaxing assumptions on the geometry.  

In this work, we address this question by focusing on conformally-coupled scalar fields in the first-order formalism of gravity. To this end, we study a one-parameter extension of conformal transformations by considering that the metric and affine properties of the manifold are independent~\cite{Nieh:1981xk,Chakrabarty:2018ybk,Izaurieta:2020kuy}---see Eq.~\eqref{conformaltransformations} below. Conformal symmetry is relevant in this context because, in four dimensions, conformally invariant metric theories are known to be finite for any asymptotically AdS solutions~\cite{Grumiller:2013mxa}; a result that can be extended for Einstein-AdS spaces in six dimensions~\cite{Anastasiou:2021tlv,Anastasiou:2022ljq,Anastasiou:2023oro} and in the presence of conformally-coupled scalar fields in four dimensions~\cite{Anastasiou:2022wjq}.\footnote{Indeed, one can embed Einstein-AdS gravity in conformal gravity by imposing Neumann boundary conditions in the Fefferman-Graham expansion~\cite{Maldacena:2011mk,Anastasiou:2016jix}.} We construct a scalar-tensor theory in the first-order formalism of gravity that remains invariant under these transformations and obtain the field equations. By assuming a static ansatz, we solve the field equations analytically and obtain an asymptotically AdS black hole solution with nontrivial torsion sourced by the scalar field. We analyze different properties of the solution putting particular attention on the torsional singularities. Remarkably, the scalar field becomes fully regular, while the curvature and torsion singularities coalesce at the origin of the spacetime. The solution is continuously connected to previous results reported in the literature in the limit where torsion vanishes. In contrast to previous findings, this limit can be achieved for a particular value of the conformal parameter without trivializing the scalar field. Finally, we analyze the asymptotic behavior of the solution and provide evidence that the value of the conserved charges will change due to the presence of torsion.

The manuscript is organized as follows: In Sec.~\ref{sec:theory}, the one-parameter extension of conformal transformations is discussed and a scalar-tensor theory that remains invariant under the latter is proposed. In Sec.~\ref{sec:solution}, the solution is presented by assuming a static ansatz with a compact constant-curvature transverse section. Section~\ref{sec:properties} is devoted to analyzing the physical properties of the black hole solution, focusing on the torsional effects on the geometry. Finally, in Sec.~\ref{sec:discussion} we present a summary and discussion. Appendix~\ref{sec:appendix} is included for the sake of comparison where we rewrite the action and field equations with tensor components.

\section{Conformally coupled scalar fields in first-order gravity}\label{sec:theory}
	
Here, we discuss the dynamics and the symmetries of the theory we are interested in. To this end, we focus on a scalar-tensor theory that remains invariant under a one-parameter family of conformal transformations. In particular, we consider the metric and connection as independent fields; this is usually regarded as the first-order formalism of gravity, where torsion is not assumed to vanish beforehand. This allows one to extend the typical conformal scalar-tensor couplings in (pseudo)-Riemannian geometries to theories in which torsional degrees of freedom are present.  

Let $\mathcal{M}_4$ be a four-dimensional Lorentzian manifold endowed with a metric tensor $g_{\mu\nu}$. Henceforth, Greek indices denote tensor components in the coordinate basis while lowercase Latin indices are used for the Lorentz orthonormal basis. We denote by $\Omega^{p}(\mathcal{M}_4)$ the set of $p$-forms defined over $\mathcal{M}_4$. Change of frame matrix components $e^{a}{}_{\mu}$ help us to define the tetrad $1$-form $e^{a}=e^{a}{}_{\mu}\mathrm{d}x^{\mu}$.\footnote{From these definitions, one can define a Lorentz-vector basis $E_a = E^{\mu}{}_a\partial_\mu$, such that it is dual to the tetrad $1$-form, that is, $e^{a}{}_\mu E^{\mu}{}_b = \delta^a_b$ and $e^{a}{}_\mu E^{\nu}{}_a = \delta^\nu_\mu$.} The line element is given by
\begin{equation}
\mathrm{d}s^2=g_{\mu\nu}\mathrm{d}x^{\mu}\otimes\mathrm{d}x^{\nu}=\eta_{ab}e^{a}\otimes e^{b}\,,
	\end{equation}
where $\eta_{ab}=\mathrm{diag}(-1,1,1,1)$ denotes the Minkowski metric and, consequently, we have the local mapping $g_{\mu\nu}(x)=e^{a}{}_{\mu}(x)e^{b}{}_{\nu}(x)\eta_{ab}$.
Since we are considering a Riemann-Cartan geometry, we introduce the Lorentz connection $1$-form, $\omega^{ab}=\omega^{ab}{}_{\mu}\mathrm{d}x^{\mu}$, as an independent field. From these quantities, one defines the curvature and torsion $2$-forms which are given by
\begin{align}
R^{ab}&=\mathrm{d}\omega^{ab}+\omega^{a}{}_{c}\wedge \omega^{cb}\,, \label{riem}\\
T^{a}&=\mathrm{d}e^{a}+\omega^{a}{}_{b}\wedge e^b\,,
\end{align}
respectively. They satisfy the Bianchi identities $\Diff{R^{ab}=0}$ and $\Diff{T^a}=R^{a}{}_b\wedge e^b$. Additionally, the Lorentz connection can always be decomposed into their torsion-free and contorsional pieces, that is, $\omega^{ab}=\mathring{\omega}^{ab} + K^{ab}$, where the Levi-Civita connection satisfies the torsion-free condition $\diff{e^a}+\mathring{\omega}^{a}{}_b\wedge e^b=0$ and $K^{ab}=K^{ab}{}_{\mu}\diff{x^\mu}$ is the contorsion $1$-form defined via $T^a=K^{a}{}_b\wedge e^b$.

We consider a scalar-tensor theory where the gravitational dynamics for the tetrad, Lorentz connection, and scalar field is dictated by the action functional
\begin{align}
I[e,\omega,\phi]&=\int_{\mathcal{M}_4} \Bigg[\frac{1}{4\kappa}\epsilon_{abcd}\left(R^{ab}-\frac{\Lambda}{6} e^{a}\wedge e^{b}\right)\wedge e^{c}\wedge
	e^{d} \notag \\
 \label{exact}
	&-\frac{1}{24}\epsilon_{abcd}\left( \phi^2R^{ab} + \left\{
	\lambda\left[ 1- \frac{\lambda}{2} \right]  Z^{2}+V(\phi)\right\}e^{a}\wedge e^{b} +4\lambda \phi Z^{a} T^{b}\right)\wedge e^{c}\wedge e^{d}\Bigg]\,.
\end{align}
where $Z^a = e^{a}{}_\mu\nabla^\mu\phi$. Here, $\kappa=8\pi G_N$ is the gravitational constant, $\Lambda$ is the cosmological constant, $V(\phi)$ is a potential for the scalar field $\phi$, and $\lambda$ is a dimensionless parameter that characterizes the extended Weyl transformation for the Lorentz connection. Additionally, the kinetic term of the scalar field is constructed in terms of $Z^2=Z_aZ^a$. Neglecting the potential $V(\phi)$, the scalar-tensor sector of the action~\eqref{exact} remains invariant under the one-parameter family of Weyl transformations
\begin{subequations}\label{conformaltransformations}
    \begin{eqnarray}
	e^a \to& \bar{e}^{a}  & = \exp[\sigma(x)] e^{a}\label{cf1}\,,\\
	\omega^{ab} \to& \bar{\omega}^{ab}  &= \omega^{ab} + \lambda \,\theta^{ab}\label{cf2}\,, \\
    \phi \to& \bar{\phi} &= \exp[-\sigma(x)]\phi\,,
\end{eqnarray}
\end{subequations}
where $\theta^{ab} = 2e^{[a}e^{b]}{}_\mu\nabla^\mu\sigma$ and $0<\lambda<1$. This, in turn, implies that the contorsion $1$-form transforms as $K^{ab}\to \bar{K}^{ab}=K^{ab}+(\lambda-1)  \theta^{ab}$. Notice that, if $\lambda\to0$, the Lorentz connection remains invariant under Weyl rescalings~\cite{Nieh:1981xk} while, if $\lambda\to1$, the contorsion does~\cite{Chakrabarty:2018ybk,Izaurieta:2020kuy}. We refer to these two limits as the exotic and canonical Weyl rescalings, respectively. 

The field equations are obtained by performing arbitrary variations with respect to the tetrad, Lorentz connection, and the scalar field, giving
\begin{subequations}\label{eom}
    \begin{align}
 0&=	\frac{1}{2}\epsilon_{abcd}R^{ab}\wedge e^c -\frac{\Lambda}{3!} \epsilon_{abcd}e^{a}\wedge e^{b}\wedge e^{c}-\kappa \tau_d \,,  \label{eomm}   \\
0&=		\epsilon_{abcd}T^{c}\wedge e^{d} -\kappa\sigma_{ab}\,,\label{eomt} \\
0&= \epsilon_{abcd}\left[\lambda(2-\lambda) \text{D} Z^a \wedge e^b-\frac{1}{2}\phi R^{ab}- \lambda(3\lambda-5)Z^a\wedge T^b+\lambda \text{d}\phi \wedge \text{I}^a(T^b)\right]\wedge e^c \wedge e^d \notag\\
	&+\epsilon_{abcd}\left[\lambda\phi \text{D}(\text{I}^a T^b)\wedge e^c -2\lambda \phi\text{I}^a(T^b)\wedge T^c-\frac{1}{4} \frac{\mathrm{d} V(\phi)}{\mathrm{d}\phi} e^a \wedge e^b \wedge e^c \right]\wedge e^d  \,, \label{eomphi}
\end{align}
\end{subequations}
respectively. Throughout this manuscript, $\text{I}^a$ denotes the inner contraction operator along the dual vector basis to the tetrad $1$-form\footnote{See Ref.{~\cite[Section 3.2]{Barrientos:2019msu}} for further details.}. Additionally, we have defined the stress-energy and spin density $3$-forms as $\tau_a$ and $\sigma_{ab}$, respectively; they are
\begin{align}
\tau_d &= \frac{1}{3}\epsilon_{abcd}\left(\frac{\phi^2}{4}R^{ab} + \lambda\phi Z^a T^b \right)\wedge e^c + \frac{1}{6}\epsilon_{abcd}\left[\frac{\lambda(\lambda-2)}{2}Z^2+V(\phi) \right]e^a\wedge e^b\wedge e^c \notag \\
 &- \frac{\lambda}{6}\epsilon_{abcd}\left[\phi\Diff{Z^a}+(3\lambda-5)Z^a\,\diff{\phi} + \phi Z_n\text{I}^nT^a \right]\wedge e^b\wedge e^c\,, \\
 \sigma_{ab}  &=\frac{\left(  1-\lambda\right) }%
	{6-\kappa\phi^{2}}\epsilon_{abcd}\, \phi\,\diff{\phi}\wedge e^{c}\wedge e^{d}\,.
\end{align}
The field equation for the connection can be solved algebraically for the torsion in terms of the scalar field and derivatives thereof. The solution is given by
\begin{equation}
T^{a}=\frac{\kappa\left(  1-\lambda\right)  }{\left( 6-\kappa
\phi^{2}\right)}\,\phi\,\diff{\phi}\wedge e^{a}\,,   \label{torsol}
\end{equation}
where $\phi^2\neq\tfrac{6}{\kappa}$. Thus, we conclude that the nonminimal coupling of the scalar field sources the nontrivial torsion in this theory. This has been observed in Refs.~\cite{Toloza:2013wi,Cid:2017wtf,Barrientos:2017utp,Barrientos:2019awg} as well. Moreover, in the limit $\lambda\to1$, the torsion vanishes independent of the value of the scalar field. In contrast, if $\lambda\neq1$, the torsion is nontrivial as long as the scalar field is not constant. In what follows, we solve the remaining field equations by assuming a static ansatz and show that the system admits a black hole solution with nontrivial torsion.

\section{Torsional black holes dressed with scalar fields\label{sec:solution}}

In this section, we explore the space of solutions of the theory by solving the field equations~\eqref{eom}. To do so, we consider a one-parameter extension of the standard quartic potential, that is,
\begin{equation}\label{Vphi}
		V(\phi)=\frac{\Lambda\pi\,G\,J^2(\phi)}{9} \left[ \left(\frac{3+\sqrt{12\pi G}\phi}{3-\sqrt{12\pi G}\phi}\right)^{-\sqrt{\lambda(2-\lambda)}}+\left(\frac{3+\sqrt{12\pi G}\phi}{3-\sqrt{12\pi G}\phi}\right)^{\sqrt{\lambda(2-\lambda)}} \right]
-\frac{\Lambda}{8\pi G}\,,
	\end{equation}
 where $J(\phi)=\phi^2 - \frac{3}{4\pi G}$. The conformal quartic potential is obtained in the vanishing-torsion limit, that is, $\lambda\to1$. It is worth noticing that Eq.~\eqref{Vphi} has extrema at $\phi_0=0$ and at $\phi_0=\pm\sqrt{\frac{3}{4\pi G}}$. The latter, however, represents a torsional singularity as one can check from Eq.~\eqref{torsol}. Indeed, we will see that these points correspond to limiting values of the scalar field. Stability of the latter will depend on the value of the cosmological constant at these points. 
The potential satisfies $V''(0)=-\tfrac{2\Lambda}{3}(\lambda-1)^2 $.\footnote{For $\lambda\neq 1$ this potential contributes to the mass term of the scalar field.} Thus, if $\Lambda<0$ ($\Lambda>0$), the extremum at $\phi_0=0$ represents a global minimum (maximum). It should be noticed that, for $\lambda=1$, the torsion vanishes and we recover the results obtained in Ref.~\cite{Martinez:2004nb}.

We assume a static metric whose codimension-2 hypersurfaces of constant $t-r$ represent locally a constant curvature space. In particular, we consider 
\begin{align}
    \diff{s^2} = h(r)\left(-f(r)\diff{t^2} + \frac{\diff{r^2}}{f(r)} + r^2\diff{\Sigma_{(k)}}^2 \right)\,,\;\;\;\;\; \mbox{where} \;\;\;\;\; \diff{\Sigma_{(k)}^2} =\frac{\diff{\vec{x}}\cdot\diff{\vec{x}}}{\left(1+\frac{k}{4}\vec{x}\cdot\vec{x} \right)^2}
\end{align}
represents the line element of a compact transverse section of constant curvature $k$ and local coordinates $\vec{x}=(x^1,x^2)$, with $k=\pm1,0$.
Additionally, the scalar field compatible with the isometries of this metric depends on the radial coordinate only, namely, $\phi=\phi(r)$.

As we mentioned above, the field equation for the connection can be solved for the torsion in terms of the scalar field and derivatives thereof, whose solution is given in Eq.~\eqref{torsol}. This, in turn, implies that the functions $\omega_I(r)$, with $I=1,\ldots,8$, can be solved algebraically in terms of the scalar field and the metric functions. The nontrivial pieces of the connection are found to be
\begin{subequations}
    \begin{align}
     \omega_1(r) &= \frac{1}{h(r)}\left[\sqrt{h(r)f(r)}\right]' + \frac{(1-\lambda)\sqrt{f(r)}\,\phi(r)\phi'(r)}{J(\phi)\sqrt{h(r)}}\,, \\
     \omega_5(r) &= - \frac{\sqrt{f(r)}}{2r^2h^{3/2}(r)}\left[h(r)\,r^2 \right]' - \frac{(1-\lambda)\sqrt{f(r)}\,\phi(r)\phi'(r)}{J(\phi)\sqrt{h(r)}}\,,
 \end{align}
\end{subequations}
 where prime denotes differentiation with respect to the radial coordinate, i.e. $'=\diff{}/\diff{r}$, and the other components of the connection vanish on shell. The field equations for the tetrad and the scalar field are solved by 
 \begin{subequations}\label{sol}
     \begin{align}
     f(r) &= k\left(1+\frac{\mu\,G}{r} \right)^2 - \frac{\Lambda r^2}{3}\,, \\
     h(r) &= \frac{r(r+2\mu G)}{4(r+\mu G)^2}\left[2+\left(1+\frac{2\mu G}{r} \right)^{\frac{1}{\sqrt{\lambda(2-\lambda)}}} + \left(1+\frac{2\mu G}{r} \right)^{-\frac{1}{\sqrt{\lambda(2-\lambda)}}} \right]\,, \\
     \phi(r)&= -\sqrt{\frac{3}{4\pi G}}\left[\frac{1-\left(1+\frac{2\mu G}{r} \right)^{\frac{1}{\sqrt{\lambda(2-\lambda)}}}}{1+\left(1+\frac{2\mu G}{r} \right)^{\frac{1}{\sqrt{\lambda(2-\lambda)}}}}\right]\,,
 \end{align}
 \end{subequations}
 where $\mu$ is an integration constant. Notice that the solution in Eq.~\eqref{sol} is continuously connected to that of Refs.~\cite{Martinez:2004nb,Martinez:2005di} in the limit $\lambda\to1$. Indeed, the causal structure of this configuration is the same as the one in those references. If $\mu>0$ for $r\in\mathbb{R}_{>0}$, then the scalar field is bounded as $0<\phi< \sqrt{\frac{3}{4\pi G}}$. Conversely, if $\mu<0$, reality on the scalar field implies that the range of the radial coordinate is $r>-2 \mu G$ while the scalar field is bounded according to $-\sqrt{\frac{3}{4\pi G}}<\phi< 0$. Thus, since the scalar field is constant as $r\to0$, we conclude that it is real and everywhere regular if $(2\lambda-\lambda^2)^{-1/2}\in \mathbb{N}_{>1}$.  

 \section{Properties of the solution\label{sec:properties}}

 Let us discuss the most relevant properties of the solution. First, the asymptotic behavior of the metric is modified by the presence of torsion when $\lambda\neq1$, since
      \begin{align}
     F(r) &\approx \frac{\Lambda r^2}{3} - k - \frac{\Lambda\mu^2G^2(\lambda-1)^2}{3\lambda(\lambda-2)} - \frac{2\mu G}{r}\left(k - \frac{\Lambda \mu^2G^2(\lambda-1)^2}{3\lambda(\lambda-2)} \right) + \mathcal{O}(r^{-2})\,, \\
     \label{falloff}
     H(r) &\approx -\frac{\Lambda r^2}{3}+ k - \frac{\Lambda\mu^2G^2(\lambda-1)^2}{3\lambda(\lambda-2)} + \frac{2\mu G}{r}\left(k+ \frac{\Lambda\mu^2G^2(\lambda-1)^2}{3\lambda(\lambda-2)} \right) + \mathcal{O}(r^{-2}) \,,
 \end{align}
 where we have defined $F(r):=h(r)f(r)$ and $H(r):=h^{-1}(r)f(r)$. Notice that the presence of torsion generates an effective curvature of the transverse section as one can see from the zeroth-order term in the radial asymptotic expansion of $g_{tt}$. Additionally, we expect that this behavior will change the value of the conserved charges as the torsion modifies the value of the $\mathcal{O}(r^{-1})$ term as $r\to\infty$. Since $0<\lambda<1$, one can see that the parameter $\mu$ can be negative (positive) for certain values of $\Lambda>0$ ($\Lambda<0$).

This solution has a curvature singularity when $r\to0$ as it can be seen by computing its Kretschmann invariant. However, it is hidden behind a horizon at $r=r_h$ defined by the largest positive root of the polynomial $f(r_h)=0$. Remarkably, if $\mu>0$, the presence of torsion renders the scalar field fully regular, in contrast to their Riemannian counterpart which develops a scalar's singularity for a finite value of the radial coordinate. Additionally, there exists a torsional invariant that can be computed from $T=T_{\mu\nu\lambda}T^{\mu\nu\lambda}$. Evaluating the latter on the solution of Eq.~\eqref{sol}, we obtain
\begin{align}\label{Torinv}
    T &=\frac{8G\mu^2  (a^2 -1) \left(\frac{G \mu}{r} +1\right)^2 \left(\frac{2 G \mu}{r}+1\right)^a \left[\left(\frac{2 G \mu}{r}+1\right)^a-1\right]^2 \left[3 k \left(\frac{G \mu}{r} +1\right)^2-\Lambda r^2\right]}{\pi  r^4 \left(\frac{2 G \mu}{r} +1\right)^3 \left[\left(\frac{2 G \mu}{r}+1\right)^a+1\right]^4}\,,
\end{align}
where $a=(2\lambda-\lambda^2)^{-1/2}$ and, since $\lambda \in (0,1]$, we have that $a>1$. 

In the case when $\Lambda=0$, the potential vanishes identically. From Eq.~\eqref{sol}, one notices that the existence of a horizon demands that $k=1$ and $\mu<0$. Thus, the topology of the transverse section is that of $\mathbb{S}^2$. In the limit $\lambda\to1$, this configuration reduces to the BBMB solution found in Refs.~\cite{Bekenstein:1974sf,Bocharova:1970skc}. If $0<\lambda<1$, however, the scalar field is regular at the horizon located at $r=-\mu G$ due to the presence of torsion, in contrast to the aforementioned solution. Nevertheless, an inspection of Eq.~\eqref{Torinv} reveals that there exists a torsional singularity at $r=-2\mu G$ which lies outside of the horizon if $1<a<3$. Therefore, this case represents a naked torsional singularity.

Let us focus on the case with negative cosmological constant $\Lambda=-3 \ell^{-2}$. First, notice that the existence of event horizons demands that $k=-1$. In this case, the topology of the horizon is $\mathbb{H}^2/\Gamma$, where $\Gamma$ is a discrete subgroup of $SO(2,1)$ such that the transverse section has finite volume. Then, there are two possible solutions depending on the sign of $\mu$. 
For $\mu>0$, hairy torsional black hole possess a single event horizon located at  
\begin{equation}\label{Horizon}
r_{+}=\frac{\ell}{2}\left(1+\sqrt{1+\frac{4\mu G}{\ell}}\right).
\end{equation}
The torsion and scalar field are regular over this hypersurface. However, a curvature and torsional singularity occur at the origin $r=0$. Since the connection is metric compatible, the causal nature of this black hole is equivalent to that of Ref.~\cite{Martinez:2005di}. The Hawking temperature in this case is given
\begin{align}
    T_H = \frac{\sqrt{F'(r)H'(r)}}{4\pi}\Bigg|_{r=r_+} = \frac{1}{2\pi\ell}\left(\frac{2r_+}{\ell} - 1 \right)\,,
\end{align}
which is positive definite by virtue of Eq.~\eqref{Horizon} and, recall, $F(r)$ and $H(r)$ have been defined below Eq.~\eqref{falloff}. For $\mu=0$, the scalar field vanishes as well as the torsion and the metric becomes global AdS. If $\mu<0$, the absence of naked singularities implies that the integration constant must be bounded according to $\mu\geq-\tfrac{\ell}{4G}$. Then, the solution has three horizons given by
\begin{subequations}
\begin{align}
    r_{--} &= \frac{\ell}{2}\left(-1+\sqrt{1-\frac{4\mu G}{\ell}} \right) \,, \\
    r_- &= \frac{\ell}{2}\left(1-\sqrt{1+\frac{4\mu G}{\ell}} \right) \,, \\
    r_+ &= \frac{\ell}{2}\left(1+\sqrt{1+\frac{4\mu G}{\ell}} \right) \,.
\end{align}    
\end{subequations}
Additionally, reality conditions on the torsion and scalar field demand that $r \geq -2 G \mu$. Then, the latter condition can be combined with bound $\mu\geq-\tfrac{\ell}{4G}$ to give $r\geq \ell/2$. Indeed, one can see that a torsional singularity occurs at $r=\ell/2$ if $1<a<3$ [cf. Eq.~~\eqref{Torinv}]. This, in turn, implies that the autoparallels could end at the torsional singularity, before reaching the curvature singularity. Nevertheless, the torsional singularity lies behind the event horizon, namely, $r_{+}>\ell/2$, where $r_{+}$ is given in Eq.~\eqref{Horizon} and, therefore, it does not represent a naked singularity.

In the case of a positive cosmological constant, say $\Lambda=3/\ell^2$, with $\ell$ being the de Sitter radius, the absence of naked singularities demands that $k=1$. If $\mu>0$, there is a unique cosmological horizon located at $r_+$ where the anti-de Sitter radius must be replaced by the de Sitter one in Eq.~\eqref{Horizon}. Therefore, the lack of an event horizon implies that this case represents a naked singularity. If $\mu=0$, the scalar field vanishes and the metric is that of global de Sitter space. If $\mu<0$, however, there is a torsional singularity at $r=\ell/2$ which lies between the event and the cosmological horizon if $1<a<3$. Thus, this case is excluded by the cosmic censorship conjecture. Finally, in the extremal case, namely, if $\mu=-\tfrac{\ell}{4G}$, the event and cosmological horizons coalesce into a single horizon. However, the torsion singularity is located at this single horizon, rendering the latter a naked singularity. Therefore, in the presence of torsion, we conclude that only the asymptotically anti-de Sitter black hole is admissible.

It is well known that the notion of geodesics and autoparallels do not necessarily coincide if torsion is present. In the case of geodesics, the Killing theorem implies that, if $u=u^\mu\partial_\mu$ is tangent to a geodesic, i.e. $u^\mu\mathring{\nabla}_\mu u^\nu=0$, and $\xi=\xi^\mu\partial_\mu$ is a Killing vector, then the product $u^\mu\xi_\mu$ is constant along the geodesic. As discussed in Ref.~\cite{Peterson:2019uzn}, this conservation law can be extended in the case of autoparallels with tangent $v=v^\mu\partial_\mu$, i.e. $v^\mu\nabla_\mu v^\nu = 0$, by introducing a new notion of Killing vectors such that they satisfy $\nabla_{(\mu}\xi_{\nu)}=0$, where $\nabla$ is the torsionful connection. These are referred to as T-Killing vectors. The existence of the latter can be analyzed by solving the T-Killing equation, which can be written explicitly in terms of the contorsion as
\begin{equation}\label{T-KillingEq}
    \mathring{\nabla}_{(\mu}\xi_{\nu)}-{K^\lambda}_{(\mu \nu)}\xi_{\lambda}=0\,.
\end{equation}
Using the definition of the contorsion $1$-form $K^{ab}$ given in Sec.~\ref{sec:theory} and the solution of the torsion $2$-form in Eq.~\eqref{torsol}, we find that, projection onto the spacetime components of the former, that is, $K^{\alpha\beta}{}_\mu = E^{\alpha}{}_a E^{\beta}{}_b K^{ab}{}_\mu$, yields
\begin{equation}
    K_{\rho\mu \nu} = \frac{2\kappa (\lambda-1) \phi}{6-\kappa \phi^2}g_{\nu[\rho}\partial_{\mu]}\phi  \,.
\end{equation}
Assuming an ansatz for the time-like T-Killing vector of the form $\xi=A(r)\partial_t$, then Eq.~\eqref{T-KillingEq} becomes
\begin{equation}
    A'(r)+\frac{\kappa (\lambda-1)  \phi \phi'}{6-\kappa\phi^2}A(r)=0\,,
\end{equation}
where $\phi=\phi(r)$. The general solution to this equation is
\begin{equation}
    A(r)= A_0 \left(6-\kappa \phi^2\right){}^{\frac{\lambda-1}{2}}\,,
\end{equation}
with $A_0$ being an integration constant. Notice that, since $\phi^2\neq\tfrac{6}{\kappa}$, this vector is nondegenerate. Moreover, in the limit of vanishing torsion, i.e. $\lambda\to1$, the standard time-like Killing vector that generates the temporal isometries is recovered. Indeed, the norm of this T-Killing vector is given by
\begin{align}
    \xi\cdot\xi = -h(r)f(r)A^2(r)\,.
\end{align}
Thus, since $\xi$ is nondegenerate, we see that it becomes null at the horizon. Therefore, we conclude that the horizon defined at Eq.~\eqref{Horizon} is also a T-Killing horizon.

\section{Discussion\label{sec:discussion}}

In this work, we consider a one-parameter family of conformal transformations in the first-order formalism of gravity. To accomplish this, we exploit the fact that the tetrad and spin connection are regarded as independent fields in this setup. Then, by considering the standard conformal weight for a scalar field, we construct a conformal coupling in the presence of torsion. The dynamics is dictated by the Einstein-Cartan term and the scalar-tensor conformal coupling alongside a scalar potential (\ref{eom}). The nonminimal coupling of the scalar field induces a nontrivial torsion. Remarkably, we find that there is a particular value for the parameter $\lambda$ in Eq.(\ref{cf2}) that sets the torsion to zero without trivializing the scalar field. Indeed, this particular point leaves the torsion invariant under conformal transformations, while the metric and the scalar field transform in the standard way.    

To look for black hole solutions, we assume a static ansatz for the metric, connection, and scalar field. We solve the field equations analytically in the presence of a one-parameter extension of the quartic scalar potential that becomes conformal in the limit $\lambda\to1$. We obtain an asymptotically AdS black hole solution with a compact horizon of negative curvature and nontrivial torsion dressed with scalar fields. In the vanishing torsion limit, the solution is continuously connected to that of Refs.~\cite{Martinez:2002ru,Martinez:2005di}. Remarkably, the torsion renders the scalar field everywhere regular, in contrast to the black hole solution in which the torsion vanishes. Nevertheless, there appears a torsional singularity that lies behind the event horizon, such that there is no violation of the cosmic censorship conjecture. We solve the T-Killing equation~\cite{Peterson:2019uzn} and conclude that the event horizon is also a T-Killing horizon.

Interesting questions remain open. First, an asymptotic analysis of the solution provides evidence that the presence of torsion would modify the asymptotic charges. The computation of the latter is certainly relevant for understanding how torsion modifies the global properties of the solution. A deeper analysis of the latter will provide a starting point to study the thermodynamic properties of the solution. In fact, there have been some approaches for the calculation of black hole entropy in other theories with nontrivial torsion~\cite{Cvetkovic:2022qpt,Blagojevic:2022etm,Aviles:2023igk}. Motivated by these results, it would be interesting to obtain the free energy to first order in the saddle-point approximation and determine whether the system develops a phase transition between a maximally symmetric space and a configuration with non-vanishing torsion. This would imply that, above a certain critical temperature, the torsional configuration would be thermodynamically preferred. We postpone a detailed analysis of these questions for a future contribution.

\begin{acknowledgments}
We thank Giorgos Anastasiou, Eloy Ayón-Beato, Yuri Bonder, Branislav Cvetkovic, Daniel Flores-Alfonso, Oscar Fuentealba, Mokhtar Hassaine, Diego Hidalgo and Marcela Lagos for insightful comments and remarks. The work of C.C. is partially supported by Agencia Nacional de Investigaci\'{o}n y Desarrollo (ANID) through Fondecyt grants N$^{\mathrm{o}}$1240043, 1240048, 1210500, and 1230112. L.A. is supported by Fondecyt grant N$^{\mathrm{o}}$3220805. O.V. acknowledges Fondecyt grant N$^{\mathrm{o}}$11200742. F.I. acknowledges financial support from the Chilean government through Fondecyt grants N$^{\mathrm{o}}$1150719, 1180681 and 1211219. 
	\end{acknowledgments}

\appendix
 
\section{Action and field equations in tensor components\label{sec:appendix}}

For the sake of comparison, here we provide the action and field equations in tensor components to undertake the analysis on a coordinate basis. In this case, the action principle in Eq.~\eqref{exact} can be rewritten as
\begin{align}
I  & =\int_{\mathcal{M}_{4}}\diff{^4}x\sqrt{|g|}\left[  \frac{1}{2\kappa}\left(  R-2\Lambda\right)
-\frac{1}{12}\phi^{2}R-\lambda\left[  1-\frac{\lambda}{2}\right]
(\nabla\phi)^2-\frac{\lambda}{3}\phi T_\lambda\nabla^{\lambda
}\phi -V\left(  \phi\right)  \right]\,, 
\end{align}
where $T_\lambda := T^{\sigma}{}_{\lambda\sigma}$ is the trace of the torsion tensor. The field equation for the tetrad can be written as
\begin{equation}
G_{\mu\nu}+g_{\mu\nu}\Lambda=\kappa\, \tau_{\mu\nu}\,,\label{einstt}
\end{equation}
where $G_{\mu\nu}$ denotes the Einstein tensor constructed out of the torsionful connection and
\begin{align}
\tau_{\mu\nu}  & =\frac{1}{6}\phi^{2}G_{\mu\nu}-g_{\mu\nu}V\left(  \phi\right)
-\frac{1}{2}\lambda\left(  \lambda-2\right)  g_{\mu\nu}(\nabla\phi)^2+\frac{1}{3}\lambda\phi\nabla_{\mu}\phi
T_\nu \nonumber \\
& +\frac{1}{3}\lambda\left(  3\lambda-5\right)\left[  g_{\mu\nu}%
(\nabla\phi)^2-\nabla_{\mu
}\phi\nabla_{\nu}\phi\right]  +\frac{1}{3}\lambda\phi\left(  g_{\mu\nu}\Box\phi-\nabla_{\nu}\nabla_{\mu}\phi\right) \,,
\end{align}
is the stress-energy tensor for the conformally-coupled scalar field. Here, $\Box=\nabla_\mu\nabla^\mu$ is constructed out of the torsionful covariant derivative. It is worth noticing that the field equation Eq.~\eqref{einstt} is not symmetric in general since $\left[\nabla_\mu,\nabla_\nu \right]\phi =  T^{\lambda}{}_{\mu\nu}\nabla_\lambda\phi$. The skew-symmetric piece arises from the fact that the tetrad is not necessarily symmetric in its two indices; the antisymmetric components are related to the presence of torsion. 

The field equation for the scalar field is given by
\begin{equation}
2\lambda\left(
1-\frac{\lambda}{2}\right)  \Box\phi-\frac{1}{6}\phi R-\frac{\partial V}{\partial\phi}-\lambda\left(
\lambda-2\right)  T^\mu\nabla_\mu\phi+\frac{1}{3}\lambda
\phi\nabla_\mu T^\mu+\frac{1}{3}\lambda\phi
T_\mu T^\mu=0\,,
\end{equation}
Finally, the field equation for the spin connection is given by
\begin{equation}
T_{~\alpha\beta}^{\mu} + 2\delta^\mu_{[\alpha}T_{\beta]}=-\frac{2\kappa\left(
\lambda-1\right)  }{\left(6-\kappa \phi^{2}\right)}\delta^\mu_{[\alpha}\nabla_{\beta]}\phi^2\,,
\end{equation}
whose solution is
\begin{equation}
T_{~\mu\nu}^{\rho}= \frac{\kappa\left(\lambda-1\right)  }{\left(
6-\kappa\phi^{2}\right)  }\,\delta^\rho_{[\mu}\nabla_{\nu]}\phi^2 \,.
\end{equation}

\bibliography{GB2019}

\end{document}